\documentstyle[pre,aps]{revtex}
\def\vPhi{\mbox{\boldmath$\Phi$}}
\def\vJ{\mbox{\boldmath$J$}}

\def\vx{\mbox{\boldmath$x$}}

\begin{document}
%\draft

\title{Relative Entropy: Free Energy Associated
with Equilibrium Fluctuations and
Nonequilibrium Deviations}

\author{Hong Qian}

\address{Department of Applied Mathematics\\
University of Washington, Seattle, WA 98195, U.S.A.\\
qian@amath.washington.edu} 

\date{\today}

\maketitle

\begin{abstract}
Using a one-dimensional macromolecule in aqueous solution 
as an illustration, we demonstrate that the relative 
entropy from information theory, 
$\sum_k p_k\ln\left(p_k/p_k^*\right)$, has a natural role 
in the energetics of equilibrium and 
nonequilibrium conformational fluctuations of 
the single molecule.  It is identified as 
the free energy difference associated with a fluctuating 
density in equilibrium, and is associated with the 
distribution deviate from the equilibrium in nonequilibrium 
relaxation.  This result can be generalized to any other 
isothermal macromolecular systems using the mathematical 
theories of large deviations and Markov processes, and at the 
same time provides the well-known mathematical results with 
an interesting physical interpretations.
\end{abstract}


\section{Introduction}
\label{sec:intro}

        Entropy is the most important concept in both statistical
mechanics and information theory.  In the latter \cite{SW},
entropy is a quantity associated with any discrete probability
distribution $\{p_k\}$\cite{fn1}
\begin{equation}
          S[\{p_k\}] = -\sum_k p_k \ln p_k
\label{entropy}
\end{equation}
and its generalization to continuous probability distributions
is straightforward.  In statistical mechanics, the same
Eq. \ref{entropy} gives the entropy, in units of $k_B$, for a
canonical ensemble of a molecular system at constant temperature.
This is know as Gibbs entropy.  The probability of a molecular 
configuration, say $\{x_k\}$, is related to its free energy 
$F(\{x_k\})$ according to Boltzmann's law 
$P(\{x_k\}) \propto e^{-F(\{x_k\})/k_BT}$
where $k_B$ is the Boltzmann's constant and $T$ is the
temperature in Kelvin \cite{Hu}.  It is generally accepted,
however, that the entropy in information
theory and that in statistical mechanics, though 
share the same name, are not the same. 

	This paper is to demonstrate a second connection between 
the two fields, both are based on the theory of probability.
There is another important concept in information
theory called relative entropy $H[\{p_k\}|\{p_k^*\}]$.
It is associated with probability distributions $\{p_k\}$ and
$\{p_k^*\}$:
\begin{equation}
   H[\{p_k\}|\{p_k^*\}] = \sum_k p_k
                        \ln\left(\frac{p_k}{p_k^*}\right).
\label{re}
\end{equation}
The relative entropy has many important mathematical properties,
for example, it is positive and equals zero if and only if
$p_k=p_k^*$, and furthermore, it is a convex function of
$p_k$.

        Until now, however, the relative entropy has not found
naturally a physical interpretation in statistical mechanics, despite its
important role, as a mathematical device, in the stability analysis 
of master equations \cite{Sch} and Fokker-Planck equations \cite{LM}.
In this note, we demonstrate, using two specific examples,
that the relative entropy is in fact the free energy associated
with isothermal equilibrium fluctuations and a generalized free energy
associated with transient nonequilibrium deviations.  The two 
examples we use are from the recently
developed stochastic theory of macromolecular mechanics \cite{Q2},
however, the generalization of our result is straightforward,
and in fact is known in mathematical literature \cite{LM,El}.

	The statistical mechanical system we discuss is a single
macromolecule in aqueous solution at a constant temperature $T$.
Interest on such system is motivated by the recent experimental 
studies on single biological molecules \cite{Bai,Q2}.  We point out
that such systems provide a unique type of nonequilibrium problems
in which the momentum distribution is in rapid equilibrium due 
to collisions with the solvent molecules.  Hence the nonequilibrium
problem is only for the conformational, stochastic dynamics of the
macromolecule.  A separation of the time scales for momentum 
and conformation (over-damped mechanics) is assumed, which leads 
to the Smoluchowski equation.

	The Smoluchowski approach to nonequilibrium statistical 
mechanics of single macromolecules in aqueous solution is in parallel 
and complementary to the approach based on Boltzmann equation for gases 
and liquids.  Both approaches are based on Newtonian mechanics, but
both invoke a stochastic element {\it a priori} in dealing with 
collisions \cite{Dor}.
There is an extensive literature on the nonequilibrium
statistical mechanics, including studies on relative entropy, based on 
the Boltzmann's framework \cite{Eu}.  This work illustrates the approach
based on the Smoluchowski's framework \cite{Q2}, which is simpler,
conceptually straightforward, and applicable to biomolecular 
applications.  We note that the rate of uncompensated heat in \cite{Eu}
seems to correspond to the entropy production rate in the Smoluchowski's
framework.  The precise and concrete relation between these two 
quantities remains to be established \cite{fn3}.

\section{Equilibrium Fluctuations of a Polymer Chain}
\label{sec:equilibrium}

	Let's consider a one-dimensional polymer chain with $N$ 
identical subunits.  Each subunit is an elastic element with 
free energy function (potential of mean force) $\phi(x)$.  Let 
$x_1,x_2,..., x_{N-1}$ be the junctions between successive 
subunits. We assume that the end $x_0=0$ is anchored and the end
$x_N$ is freely fluctuating.   This model is motivated by 
the mechanical studies on giant muscle protein titin \cite{QS}. 
According to Boltzmann's law, the joint probability for all 
$\{x_k\}$ is 

\begin{equation}
     P_{eq}(\{x_k\}) = Z^{-1} \exp\left[-\frac{\phi(x_1-x_0)+
\phi(x_2-x_1)+...+\phi(x_N-x_{N-1})}{k_BT}\right]
\label{prob}
\end{equation}
where 
\[    Z = \left(\int_0^{\infty} 
	\exp\left[-\frac{\phi(z)}{k_BT}\right] dz \right)^N.  \]
The partition function of the equilibrium state of such system can 
be obtained analytically using Laplace transform \cite{Ta,Ka}. 
The theory of large deviation of level-1 serves as its rigorous 
mathematical foundation \cite{El}.   

	In a laboratory, usually only $x_N$ is observable \cite{QE}.
For this case, we obtain the marginal distribution of (\ref{prob}) 
\begin{equation}
      P_{eq}(x_N) = \int_0^{\infty}...\int_0^{\infty}
      P_{eq}(\{x_i\})\ dx_1 ...  dx_{N-1}.               
\label{pend}    
\end{equation}
Combining the probability distribution in Eq. \ref{pend} with
Boltzmann's law, we obtain a free energy function for the entire
polymer under the condition that the end of the chain is at $x_N$
\[            F(x_N) = -k_BT\ln P_{eq}(x_N).             \]
In a completely parallel fashion, one can obtain the 
free energy function $F(x_i,x_j)$ from a marginal distribution
of (\ref{prob}). 

	A different type of laboratory measurements is to obtain 
the density for subunit length.  Optical spectroscopy is 
sensitive to the length of subunits, hence it provides a
measurement on the density function
\begin{equation}
  \nu(x) = \frac{1}{N}\sum_{k=1}^N \delta\left(x-x_k+x_{k-1}\right),
\end{equation} 
which is known as the empirical measure in the theory of large
deviation of level-2 \cite{El}.

	Clearly, the function $\nu(x)$ fluctuates in an equilibrium
state since $(x_k-x_{k-1})$ fluctuates.  It has an expectation 
\[   E\left[\nu(x)\right]
       = p(x) = \frac{e^{-\phi(x)/k_BT}} 	
	           {\int_0^{\infty} e^{-\phi(x)/k_BT} dx}.    \]
For large $N$, the fluctuations of $\nu(x)$ around $p(x)$ are so 
small that one seldomly considers their existence.  Nevertheless, 
there is a free energy associated with each $\nu(x)$, and we now 
show that 
\begin{equation}
     F\left[\nu(x)\right] - F\left[p(x)\right] 
    = Nk_BT \int_0^{\infty} \nu(x)\ln\left(\frac{\nu(x)}{p(x)}\right) dx.
\label{relent}
\end{equation} 
This is in fact a mathematical result for large deviation of level-2 
\cite{El}. We give only a heuristic proof below and leave the 
rigorous treatment to the mathematical literature. 

	Let's denote  
\[      \nu_n = \int_{n\delta}^{(n+1)\delta} \nu(x) dx, 
		\hspace{0.25cm} \textrm{and}
                \hspace{0.25cm}
        p_n = \int_{n\delta}^{(n+1)\delta} p(x) dx.      \]
Because all the subunits are independent, $\nu_n$ is a 
multinomial distribution
\begin{equation}
    P\left(\nu_1,\nu_2,...\nu_m\right) = \frac{N!}{(N\nu_1)!(N\nu_2)!...
	(N\nu_m)!}\ p_1^{N\nu_1} p_1^{N\nu_2} ... p_1^{N\nu_m}.
\label{pdis}
\end{equation}
According to Boltzmann's law, the free energy difference
between configuration $\{\nu_n\}$ and $\{p_n\}$ is
\begin{eqnarray*}
      F\left[\{\nu_n\}\right]-F\left[\{p_n\}\right] 
     &=& -k_BT\ln\left(\frac{P\left(\{\nu_n\}\right)}
{P\left(\{p_n\}\right)} \right) 	\\
     &\approx& Nk_BT \sum_{n=1}^m \nu_n \ln\left(\frac{\nu_n}{p_n}\right) \\
     &\rightarrow& Nk_BT\int_0^{\infty} \nu(x) \ln\left(
	\frac{\nu(x)}{p(x)} \right) dx
	\hspace{1cm} (\delta \rightarrow 0) 
\end{eqnarray*}

	Therefore, the relative entropy in Eq. \ref{relent} 
is the free energy difference between the distribution $\{\nu(x)\}$
and its (equilibrium) expectation $\{p(x)\}$.  Relative entropy
is the free energy associated with a fluctuating density at
equilibrium \cite{fn2}!

\section{Nonequilibrium Relaxation of a Polymer Chain}  

	Our second example extends the concept of free energy
beyond an equilibrium state, and reveals its central role in the
transient, isothermal, relaxation processes to equilibria.  The 
dynamic model for the polymer chain in an aqueous solution at 
constant temperature $T$ is a Smoluchowski equation \cite{Q2}:
\begin{equation}
   \frac{\partial P(\vx,t)}{\partial t} =
    \frac{k_BT}{\eta} \nabla^2 P +
   \frac{1}{\eta}\nabla\cdot\left[\nabla U(\vx)P\right]
\label{smo}
\end{equation} 
where $\eta$ is a frictional coefficient, $\vx$ =
$(x_1,x_2,...,x_N)$, $\nabla$ = 
$\left(\frac{\partial}{\partial x_1},\frac{\partial}{\partial x_2},...,
\frac{\partial}{\partial x_N}\right)$, and 
\begin{equation}
      U(\vx) = \phi(x_1-x_0) + \phi(x_2-x_1) + ... + \phi(x_N-x_{N-1}).
\end{equation}
It is easy to verify that Boltzmann's distribution $P_{eq}(\vx)$ in
Eq. \ref{prob} is the stationary solution to the Eq. \ref{smo}.  In fact, 
the steady-state solution of (\ref{smo}) defines a stationary,
time-reversible, stochastic process with equilibrium fluctuations
\cite{QQG}. 

	How is an arbitrary distribution $P(\vx)\neq P_{eq}(\vx)$
changing with time and approaching to $P_{eq}(\vx)$?  We now show 
that a free energy functional can be introduced, and it is in fact 
the relative entropy.  Let's define a $\Psi$-function
\begin{eqnarray}
    \Psi[P(\vx)] &=& \int
    \left(U(\vx)P(\vx)
                +k_BTP(\vx)\ln P(\vx)\right)d\vx   \label{fef}\\
     &=& -k_BT\ln Z + k_BT\int P(\vx)\ln
        \left(\frac{P(\vx)}{P_{eq}(\vx)} \right)d\vx.  \nonumber
\end{eqnarray}
The first term is the Helmhotz's free energy of the entire polymer 
chain in its equilibrium state, and the second term should be 
interpreted as the 
free energy difference between the arbitrary distribution 
$P(\vx)$ and the equilibrium distribution $P_{eq}(\vx)$.

	The mathematical properties of relative entropy immediately
lead to the following statement: The system reaches equilibrium if 
and only if the free energy functional (the $\Psi$-function) is 
minimized.  Furthermore, 
if $P(\vx,t)$ changes with $t$ in a transient process according to 
Eq. \ref{smo}, then 
\begin{equation}
   \dot{\Psi}[P(\vx,t)] = - \int J(\vx,t)\Phi(\vx,t)d\vx
          \le 0 
\label{frate}
\end{equation}
where
\[    \vJ(\vx,t) = -\frac{k_BT}{\eta} \nabla P(\vx,t)
	-\frac{1}{\eta} \nabla U(\vx) P(\vx,t)    \]
and
\[   \vPhi(\vx,t) = -k_BT \nabla \ln P(\vx,t)-\nabla U(\vx)    \]
are fluxes and forces, which are both zero at equilibrium \cite{Q1}.  
Furthermore, $\vJ =P \vPhi$ and Eq. \ref{frate} is related to the 
entropy production rate \cite{Q2,QQG,Q1}.  Eq. \ref{frate} immediately 
leads to a second statement: The dynamic of $P(\vx,t)$ follows a 
path of decreasing free energy.  The free energy functional in 
(\ref{fef}) is a Lyapunov function \cite{LM} for the stochastic 
dynamics of the polymer. The relative entropy is the free energy 
difference between an arbitrary distribution and the equilibrium 
distribution.  It is associated with the nonequilibrium deviation 
from the equilibrium state.   However, it is interesting to note that 
the dynamics does not follow the steepest descent of the free energy 
functional.  The interpretation and significance of this observation 
are not clear to us at present time.  We also note that Eq. \ref{frate}
corresponds to the $H$-theorem in Boltzmann's framework. 

	It is of course not a coincidence that the relative entropy 
appears as a free energy difference in both the equilibrium and 
nonequilibrium situations.  Onsager \cite{On} has pointed out that the 
force driving the nonequilibrium relaxations is in fact the same force
causing the equilibrium fluctuations to return to its mean.  Our result,
therefore, firmly relates the force to a free energy difference
in terms of the relative entropy. 

	In summary, we have shown that the relative entropy in
information theory has a natural physical meaning in equilibrium and 
nonequilibrium statistical mechanics.  It is in fact the 
free energy difference associated with the equilibrium fluctuations
of a density function, a result known in the theory of 
large deviations.  In nonequilibrium systems, it is the
free energy difference between an arbitrary distribution and 
the equilibrium distribution.  Again the latter result is 
known in the theory of Markov processes.  What is novel of
the present note is to identify the relative entropy with
the Helmholtz's free energy for isothermal systems.  
Conversely, the mathematical theorems mentioned above become an 
integral part of the statistical physics of macromolecules.


\begin{references}
\bibitem{fn2}
Mathematically speaking, both this result and that in \cite{fn1} are 
parts of Sanov's theorem on large deviations for finite alphabets 
\cite{DZ}.  Realizing the mathematical nature of and applying the 
probabilistic method to this problem, though in a less explicit manner, 
dates back to Khinchin \cite{Hu}.
\end{references}
\end{document}